\documentclass[
 aps,
 prl,
 reprint,
 superscriptaddress
]{revtex4-2}

\usepackage{graphicx}
\usepackage{dcolumn}
\usepackage{bm}
\usepackage{graphicx,amsmath,amssymb,amsthm}
\usepackage{xcolor}
\usepackage[normalem]{ulem}
\usepackage{thmtools, thm-restate}
\usepackage{epsfig}
\usepackage{epstopdf}
\usepackage{latexsym}
\usepackage{graphicx}
\usepackage{booktabs}
\usepackage{bbm}
\usepackage{color}
\usepackage{physics}
\usepackage{tensor}
\usepackage{verbatim}
\usepackage[caption=false]{subfig}
\usepackage{tikz}
\usepackage{bigints}
\usepackage{ifthen}
\usepackage{soul}
\usepackage[hidelinks]{hyperref}

\usepackage[normalem]{ulem}

\usepackage{graphicx}% Include figure files
\usepackage{dcolumn}% Align table columns on decimal point
\usepackage{bm}% bold math
\newcommand{\TeV}{\,\mathrm{TeV}}

\begin{document}

\preprint{APS/123-QED}

\title{Separating Fermion and Gauge Resonances\\with Species-Dependent Lightcones in Warped Geometry}% Force line breaks with \\

\author{Lisa Randall}
\affiliation{Department of Physics, Harvard University, Cambridge, Massachusetts 02138, USA}
\author{Marcos Riojas}
\affiliation{Department of Physics, Harvard University, Cambridge, Massachusetts 02138, USA}
\date{\today}% It is always \today, today,
 % but any date may be explicitly specified

\begin{abstract}
Extra-dimensional flavor models generally tie together the mass scales of gauge and fermion Kaluza--Klein modes. We show how species-dependent five-dimensional lightcones, which can be generated, for example, by a simple scalar profile, can separate these scales while preserving four-dimensional Lorentz invariance and any proposed extra-dimensional flavor structure. In two illustrative models on stabilized backgrounds, right-handed quark partners lie at $6.6$--$12\TeV$ with $30\TeV$ KK gluons, or a $5.5\TeV$ gluon lies below fermion partners above $10\TeV$. We compare with deconstructed and three-brane models.

\end{abstract}
\maketitle

\section{Introduction}

Extra-dimensional localization of fermions has a long history~\cite{Rubakov:1983bb,Callan:1984sa}. In flavor models, separated or warped wavefunctions can account for hierarchies in Yukawa couplings and mixing angles~\cite{Arkani-Hamed:1999ylh,Mirabelli:1999ks,Arkani-Hamed:1999gjk,Arkani-Hamed:1999pwe,Branco:2000rb,Grossman:1999ra,Gherghetta:2000qt,Huber:2000ie}. However, when flavor is explained at low, potentially accessible energy scales, this success comes with a familiar cost. The same compactification geometry that produces fermion partners also produces gauge Kaluza--Klein (KK) modes, whose nonuniversal couplings can generate dangerous flavor-changing currents~\cite{Agashe:2004cp,Casagrande:2008hr,Csaki:2008zd,Blanke:2008zb}. This ``flavor geography'' of zero modes allows the placement of the fermion zero modes along the extra dimension to set both the Yukawa couplings and the nonuniversal couplings as well. Flavor violation constraints generally require a high KK scale, and the competing demands of meson mixing, which favors larger Yukawa couplings, and dipoles, which favor smaller ones, make low-scale flavor models hard to find. At a more basic phenomenological level, models are already straining naturalness, both for the Higgs mass and for flavor. Raising the common KK scale to accommodate those constraints, and allowing for a little hierarchy \cite{Cheng:2003ju}, generally moves any new states to energies that can elude even future high-energy colliders. An important question is whether satisfying these requirements automatically precludes the presence of accessible particles.

Two ideas stand out for their attempts to evade the automatic connection among KK scales. Intermediate-brane warped models~\cite{Agashe:2016rle} add an intermediate boundary or endpoint, so that matter partners can be heavier than gauge resonances. Vectorlike-chain models~\cite{Arkani-Hamed:2026wwy} generate the hierarchy in four dimensions in such a way that these resonances simply do not appear. Here we show that a single five-dimensional interval offers a third route, which shares aspects of both.

The missing freedom we exploit here is nonuniversal propagation in the fifth direction. A field's KK gap is determined not only by the size and warping of the extra dimension, but also by the coefficient of its derivative along that direction. This coefficient fixes the local fifth-direction characteristic speed and hence the opening of the field's five-dimensional lightcone into the compact direction. If the fermion cone is narrower than the gauge-field cone, the fermion excitation energy is parametrically smaller, and the reverse applies with the opposite hierarchy. The ordinary four-dimensional lightcone remains common: only the relation between fifth-direction momentum and four-dimensional mass changes. At the two-derivative level used in Models A and B, the graviton speed is assumed unchanged, in which case  the slowed species would lie inside the gravitational cone and no superluminal  propagation would be introduced.
Species-dependent KK scales and limiting speeds have been studied in bulk Lorentz-violating theories (including warped models)~\cite{Dubovsky:2001fj,Rizzo:2005um,Rizzo:2010vu}, in aether compactification~\cite{Carroll:2008pk,Oliveira:2020chr}, and in deconstructed dimensions~\cite{Arkani-Hamed:2001kyx}. Here we incorporate this freedom into higher-dimensional flavor models. The mechanism is distinct from brane-localized kinetic terms, which alter boundary conditions and overlaps rather than the bulk fifth-direction derivative~\cite{Carena:2004zn,Davoudiasl:2002ua}.

In the framework of Ref.~\cite{Rizzo:2010vu}, Lorentz violation introduced in the gravity sector of a warped geometry gave the graviton a mass. A coupling absent at tree level would still be generated radiatively by the matter couplings, so one might expect the same obstruction here. However, that argument treated the preferred-direction coefficient as a fixed background.  In the full covariant Einstein--scalar system, the backreaction of a dynamical field removes the apparent tensor mass term, leaving the usual Sturm--Liouville equation with a constant massless graviton zero mode. Here we  generate the preferred direction specifically from dynamical scalar gradients and vary the complete action before imposing the background. We denote the dynamical bulk scalars collectively by $\phi_A$; in the models below the relevant fields are $\phi_A=(\sigma,\chi)$, with $\sigma$ supplying most of the lightcone modification and $\chi$ used primarily for stabilization. In particular, the generally covariant scalar-gradient operators that generate the species-dependent light cones for the matter fields can be written as below. After the scalar background is imposed, the same physics can be described by a species-dependent inverse vielbein,

\begin{equation}
\begin{gathered}
-\frac{i\alpha_i^{AB}}{2\Lambda_0^5}\,
 \partial_M\phi_A\,\partial^N\phi_B\,
 \bar\Psi_i\Gamma^M\overleftrightarrow D_N\Psi_i,\\
\frac{\eta_a^{AB}}{2g_{5a}^2\Lambda_0^5}\,
 \partial^M\phi_A\,\partial^N\phi_B\,
 F^a_{MP}F_N^{a\ P}.
\end{gathered}
\label{eq:covariantcones}
\end{equation}
 
\begin{equation}
 \widetilde e_{s\,a}{}^{M}=e_a{}^{M}+(v_s-1)n_a n^M,
 \qquad
 n^M=\frac{\partial^M\sigma}{\sqrt{(\partial\sigma)^2}},
\end{equation}
where the displayed $n^M$ is the  direction for the single-gradient $\sigma$ background used in the models, so the relevant nonzero component will be $n^5$. This is the precise sense in which different particle species have different five-dimensional lightcones. At the two-derivative level used for the models, the gravity sector is ordinary Einstein--scalar gravity. Even if radiative corrections had modified the graviton light speed,  the transverse--traceless sector can still be reduced to a second-order Sturm--Liouville problem, so one may write schematically
\begin{equation}
 -\partial_y\!\left(e^{-4A}\mathcal Z_5(y)\,\partial_yh_n\right)
 =m_n^2\,e^{-2A}\mathcal Z_4(y)\,h_n .
\label{eq:tensor}
\end{equation}
Thus even if $\mathcal Z_4\neq\mathcal Z_5$, so that the graviton's fifth-direction propagation speed is shifted, $h_0=\mathrm{const}$ remains an $m_0=0$ solution. A physical massless graviton additionally requires compatible covariant boundary conditions and a finite positive norm. This is the usual tensor structure of covariant scalar--gravity warped backgrounds~\cite{DeWolfe:1999cp}. Its application to the present operators and higher-derivative corrections is discussed in Ref.~\cite{Companion}.

In this paper we generate the species-dependent propagation covariantly, so that the warped theory retains a massless graviton zero mode while allowing the matter excitation scales to separate. We show how species-dependent lightcones can lead to different phenomenological conclusions for models of flavor, and we demonstrate a rescaling that keeps the zero-mode flavor structure of any existing extra-dimensional flavor model intact while altering the partner spectrum. We implement both orderings of fermion and gauge resonances on one interval, with explicit backreacted and stabilized scalar backgrounds, in two benchmark models.

In a simple five-dimensional geometry with boundary branes and scalar gradients inducing the kinetic term modifications, there is less freedom than in a true three-brane model to restrict fields to different domains, and without additional symmetries the geometry would not itself enforce a Yukawa texture as in deconstructed models. We therefore face somewhat stricter flavor constraints than those examples. We impose two-spurion correlations in Model A and common light-doublet profiles and alignment in Model B. We view these as two of many possible models that can in principle have better flavor structure. Here our focus is the possibility of separating KK mass scales.

\section{The Mechanism}
The reason flavor symmetry can readily be accommodated is that rescaling a species' fifth-direction speed together with its bulk mass leaves its wavefunctions unchanged and multiplies its KK masses by the same factor. In a locally orthonormal frame we take
\begin{multline}
\mathcal L\supset-\frac{1}{4g_5^2}\Big[F_{MN}F^{MN}
 -2(1-u_A^2)n^Mn^NF_{MP}F_N{}^P\Big]\\
 +\frac{i}{2}\bar\Psi\Gamma^\mu\overleftrightarrow D_\mu\Psi
 +\frac{i}{2}v_\Psi\bar\Psi\Gamma^5\overleftrightarrow D_5\Psi
 -M_\Psi\bar\Psi\Psi,
\label{eq:action}
\end{multline}
so that, for momenta large compared with the bulk mass, the fermion obeys $\omega_\Psi^2=\mathbf p^2+v_\Psi^2p_5^2$ and its lightcone has slope $v_\Psi$ in the fifth direction, while the gauge field has fifth-direction speed $u_A$. In the warped metric $ds^2=e^{-2A(y)}\eta_{\mu\nu}dx^\mu dx^\nu+dy^2$, the corresponding pre-EWSB mode equations are
\begin{align}
\left[v_\Psi(\partial_y-2A')+\frac{v_\Psi'}{2}+M\right]f_{L,n}
 &=m_{\Psi,n}e^A f_{R,n},\nonumber\\
\left[-v_\Psi(\partial_y-2A')-\frac{v_\Psi'}{2}+M\right]f_{R,n}
 &=m_{\Psi,n}e^A f_{L,n}.
\label{eq:fermionmodes}
\end{align}
The $v_\Psi'/2$ terms follow from the Hermitian fifth-direction Dirac operator in Eq.~\eqref{eq:action}; they are essential for a position-dependent speed. The original five-dimensional profiles are normalized with the usual warped kinetic measure. For a transverse gauge mode $a_n(y)$ with fifth-direction speed $u_a(y)$,
\begin{equation}
-\partial_y\!\left[u_a^2 e^{-2A}\partial_y a_n\right]
 =m_{A,n}^2 a_n .
\label{eq:gaugemode}
\end{equation}
 Considering first for simplicity a flat interval of length $L$ with constant $v_\Psi$, a fermion with bulk mass $M$ has
\begin{equation}
 f_0(y)\propto e^{-My/v_\Psi},\qquad
 m_n^2=M^2+v_\Psi^2\frac{n^2\pi^2}{L^2},
\label{eq:flat}
\end{equation}
while gauge modes sit at $n\pi/L$. Holding $M/v_\Psi$ fixed therefore keeps the zero-mode wavefunction, and hence the flavor geography, unchanged, while the entire massive fermion tower scales with $v_\Psi$. This is a special case of an exact relation that holds on any fixed geometry with fixed boundary conditions. Before electroweak symmetry breaking,
\begin{equation}
 \begin{gathered}
 v_\Psi(y)\to\lambda v_\Psi(y),\quad M(y)\to\lambda M(y)\\
 \Longrightarrow\quad f_n(y)\ \text{unchanged},\quad m_n\to\lambda m_n,
 \end{gathered}
\label{eq:rescaling}
\end{equation}
so every normalized wavefunction $f_n(y)$, for the zero mode and every KK level, is unchanged and every KK mass is multiplied by $\lambda$. The reason is that the mode equations depend only on $M/v_\Psi$. Likewise $u_a\to\lambda u_a$ rescales a gauge tower without changing its profiles. The Higgs couplings are overlaps of these wavefunctions with the fixed Higgs profile, so they are unchanged too, and so are the leading-order light-quark masses and mixings. After electroweak symmetry breaking the fermion mass matrix is
\begin{equation}
 \mathcal M=\mathcal M_{\rm KK}+\frac{v_{\rm EW}}{\sqrt2}\,\mathcal Y,
\label{eq:massmatrix}
\end{equation}
where $v_{\rm EW}$ is the Higgs vacuum value and $\mathcal Y$ is the matrix of Yukawa couplings between all modes. Under the rescaling $\mathcal M_{\rm KK}\to\lambda\mathcal M_{\rm KK}$ while $\mathcal Y$ is unchanged. The partner masses therefore move while the leading-order light-quark masses stay fixed, with corrections only from mixing with the KK modes. The rescaling is a choice rather than a requirement: it keeps an existing flavor geography while lowering its partner tower, but speeds and bulk masses can also be chosen independently. Changing the shape of $v_\Psi(y)$, rather than its normalization, cannot be compensated exactly with constant bulk masses. One instead retunes each constant bulk mass to restore its infrared overlap. Flat realizations of this kind, including Gaussian localization in which wall positions set the Yukawa hierarchy while a single oscillator scale sets every partner mass, are developed in the companion paper~\cite{Companion}. In a warped metric $ds^2=e^{-2A(y)}\eta_{\mu\nu}dx^\mu dx^\nu+dy^2$~\cite{Randall:1999ee} the result has a useful geometric intuition: the speed enters the mode equations through
\begin{equation}
 A_{\rm eff}(y)=A(y)-\ln v_\Psi(y),
\label{eq:Aeff}
\end{equation}
so a slower field sees its own throat, deeper by $\ln(1/v_\Psi)$. For a constant speed its KK scale is $v_\Psi$ times that of the conventional fields, while for a general profile the mass function and boundary conditions also enter. The zero-mode density,
\begin{equation}
 \rho_0(y)\propto\frac{e^{A(y)}}{v_\Psi(y)}\exp\!\Big[-2\!\int_0^y\!ds\,\frac{M}{v_\Psi(s)}\Big],
\label{eq:profile}
\end{equation}
is invariant under the rescaling of Eq.~\eqref{eq:rescaling}. A modest speed ratio of about one fifth is enough to separate the fermion and gauge scales by the same factor. As with the scalar potential, we treat this ratio as an input of the effective theory.

In a holographic reading, KK modes are resonances of a strongly coupled sector. To accommodate the interaction with the scalars $\phi_A$, we would need to include scalar operators in the holographic theory that can generate the bulk interaction
\begin{multline}
 \Delta S_i=-\frac{i}{2\Lambda_0^5}\int d^5x\sqrt{-g}\,\alpha_i^{AB}\\
 \times\partial_M\phi_A\,\partial^N\phi_B\,\bar\Psi_i\Gamma^M\overleftrightarrow{D}_N\Psi_i,
\label{eq:holoint}
\end{multline}
by choosing appropriately the boundary data associated with the interaction. The relevant term in the fermion two-point function $G_i$ is the one quadratic in the sources $j_A$,
\begin{multline}
 G_i(x,x';j)\supset\frac12\sum_{A,B}j_Aj_B\int d^4z\,d^4w\\
 \times\langle\mathcal F_i(x)\overline{\mathcal F}_i(x')\mathcal O_A(z)\mathcal O_B(w)\rangle_{0,c},
\label{eq:holoG}
\end{multline}
where $\mathcal F_i$ is the fermionic operator of the strongly coupled sector dual to $\Psi_i$, $\mathcal O_A$ is the scalar operator dual to $\phi_A$ with source $j_A$, and $\langle\cdots\rangle_{0,c}$ is a connected correlator of the undeformed sector.

The speeds arise from scalar gradients coupled to the fifth-direction kinetic terms, with a separate coefficient for each species. Operators
\begin{equation}
\begin{gathered}
-\frac{i\alpha}{2\Lambda_0^5}\,\partial_M\sigma\,\partial^N\sigma\,\bar\Psi\Gamma^M\overleftrightarrow{D}_N\Psi,\\
\frac{\eta}{2g_5^2\Lambda_0^5}\,\partial^M\sigma\,\partial^N\sigma\,F^a_{MP}F^{a\ P}_N,
\end{gathered}
\label{eq:operators}
\end{equation}
with $\Lambda_0$ the operator scale, give, on the benchmark background with $\sigma'=\nu$,
\begin{equation}
 v_\Psi=1-\alpha\frac{\nu^2}{\Lambda_0^5},\qquad u_a^2=1-\eta\frac{\nu^2}{\Lambda_0^5},
\label{eq:speeds}
\end{equation}
while leaving the four-dimensional kinetic terms canonical. The fermion speed multiplies the first-order Dirac derivative linearly, whereas the square of the gauge speed multiplies the fifth-direction gauge kinetic term. These coefficients are independent. The two examples presented below are two choices of these coefficients.

One scalar would be enough to fix the interval's length, and to slow the fermions. We choose to use two scalars so that for appropriate parameters one scalar dominantly  stabilizes the geometry and the other is primarily responsible for the modified lightcones. Both constructions are essentially equivalent in terms of the lightcone implications aside from a small $y$-dependence of the velocities from the Goldberger--Wise profile. A field $\sigma$ whose slope is constant along the interval supplies most of the slowing. A second field $\chi$ of the Goldberger--Wise type~\cite{Goldberger:1999uk} fixes the size of the interval, $kL=28$ (with $k$ the curvature scale), through a ratio of about four between its values at the two ends. This requires that $\sigma$ not fix the interval itself, which holds if the boundary terms that pin $\sigma$ at the two ends are weak enough~\cite{Companion}.

We solve the Einstein--scalar equations including the scalar backreaction on the geometry using the standard first-order scalar--gravity construction~\cite{DeWolfe:1999cp}, as a function of the boundary parameters, both with $\sigma$ alone and with the additional Goldberger--Wise field $\chi$. The scalar--metric spectrum contains the radion, the modulus associated with fluctuations of the stabilized fifth-dimensional size~\cite{Csaki:2000zn,Kofman:2004tk}. In our backgrounds the relevant mode is stable. Its mass depends on which scalar does most of the work of fixing that size. Varying the boundary parameters of the two models gives a radion mass between about $0.1$ and $1\TeV$, with the lighter values when the Goldberger--Wise field dominates and the heavier ones when $\sigma$ does~\cite{Companion}.

\begin{figure}[t]
\includegraphics[width=\columnwidth]{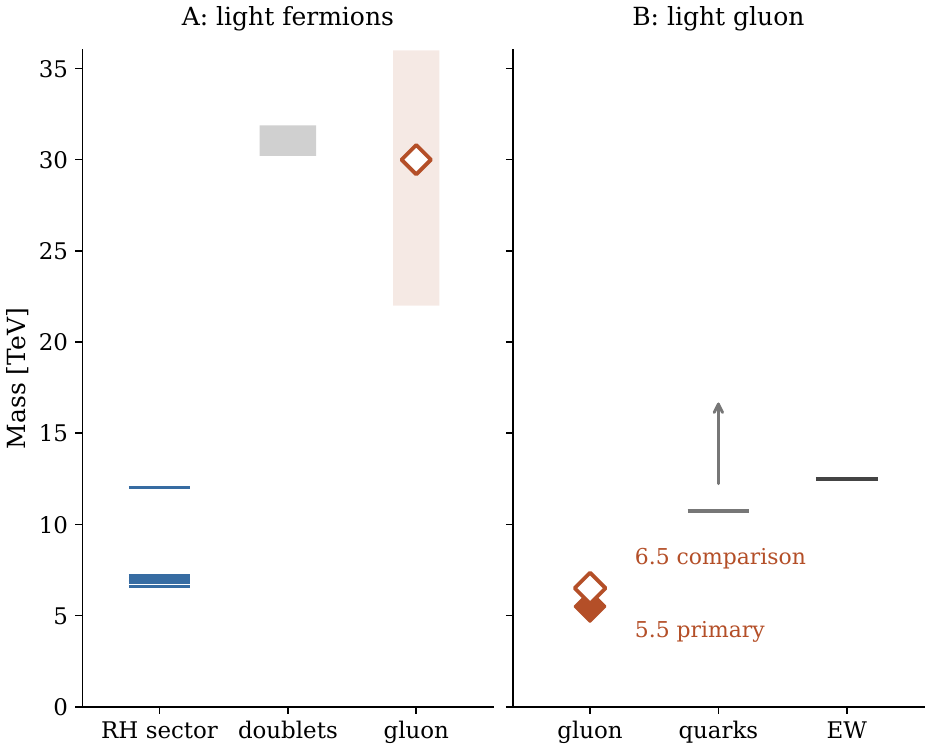}
\caption{The two models. Blue denotes the slowed right-handed fermion partners, gray denotes conventional-speed fermion or electroweak levels, and orange denotes the gluon tower. A: the blue right-handed partners lie at $6.6$--$12\TeV$, below the gray doublet first levels, while the first KK gluon has a classical eigenvalue of $30\TeV$ (orange open diamond). The pale orange band is schematic and indicates that the strongly coupled gluon's spectral weight is spread over a broad range of energies whose detailed shape is not determined. B: slowing only the gluon's fifth-direction propagation places a narrow KK gluon at $5.5\TeV$ (orange filled diamond, primary) or $6.5\TeV$ (orange open diamond, comparison), below the gray quark partners beginning at $10.74\TeV$ and electroweak KK modes at $12.5\TeV$.}
\label{fig:models}
\end{figure}

\section{Model A: Light Fermions, Heavy Gluons}
Species-dependent lightcones do not by themselves remove flavor violation, but they raise the gauge KK scale that controls the most dangerous tree-level exchange. We choose a constant-mass bulk Higgs, $M_H^2=-3k^2$, as a representative point that allows the zero-mode spectra to be matched, and keep only the Standard Model quark representations. We turn on the fermion coefficient only for the right-handed quark sector, giving a singlet speed $v_S\simeq0.19$--$0.22$ across the interval, where the modest variation comes from the position-dependent Goldberger--Wise contribution to the fermion kinetic term, while the doublets and all gauge fields keep the conventional speed. Heavy doublets suppress the chirality-changing mixing that drives dipoles. Vectorlike-chain models make the same choice for the same reason~\cite{Arkani-Hamed:2026wwy}. Flavor follows the Fitzpatrick--Perez--Randall construction in which the two Yukawa spurions also control the bulk masses~\cite{Fitzpatrick:2007sa}. The Yukawa inputs are order one, with magnitudes between $0.35$ and $2.75$ and large mixing angles before localization, while the bulk flavor tensors are correlated through the spurions.

The six first right-handed partners lie at $6.6$--$12\TeV$ (Fig.~\ref{fig:models}), three below $7\TeV$, while the doublets and first KK gluon lie near $30\TeV$. The top-singlet partner is the heaviest, at $12.019\TeV$, because its strong infrared localization raises its excitation. With tree-level exchange of the first KK gluon, the magnitude of the kaon contribution is about $1.5\%$ of the observed $|\epsilon_K|$ ($1.6\%$ when the colored KK states are included in the QCD running). The $B_d$ and $B_s$ amplitude ratios are below $0.2\%$ and $2\%$, respectively. The full contribution of the broad gluon requires low-energy matching. Without custodial protection, the total $T$ parameter is $0.027$, and the $\chi^2$ of $S$, $T$ and $U$ against the global electroweak fit, with $U$ left free, is $0.12$. The neutron electric dipole moment (EDM) obtained from the quark EDMs lies about five orders of magnitude below current limits. Complete EDM matching remains open.

The many light partners make the gluon KK resonances much broader. Even in standard warped models the KK gluon is relatively broad~\cite{Lillie:2007yh,Agashe:2006hk}. Here its strongly coupled decays into the light partners are open, and a formal sum of the lowest-order widths of all channels gives $\Gamma/M=2.30$ at $g_s^2=1.05$. This signals the breakdown of a narrow-resonance description, and the gluon is then likely too broad to appear as an identifiable resonance. Its spectrum once these strong couplings are included, and with it the size of low-energy exchange, depends on the matching to the low-energy theory. Broad gauge and fermion resonances might be an interesting feature of these models. They need not produce narrow peaks, but broad tails and interference still face collider constraints, and their virtual exchange remains, unlike in vectorlike-chain models in which gluon resonances are absent.

To measure how strongly the Higgs couples to the heavy fermion states, we use the loop-size estimate
$I_Y=2\max_q\lVert Y_h^{(q)}\rVert^2/(16\pi^2)$,
where $Y_h^{(q)}$ is the coupling matrix of the physical Higgs among the states of charge $q$ and $\lVert\cdot\rVert$ denotes its largest singular value (defined precisely in Ref.~\cite{Companion}). Including all fermion levels below $40\TeV$, it is $0.078$ for the selected bulk Higgs. An infrared-brane Higgs gives $I_Y=1.07$ at the same threshold and requires custodial protection, which is why we chose a bulk rather than a brane Higgs. The singlet-dominated partners have $W/Z/h$ widths below $0.16\%$ of their masses at lowest order and decay promptly, while the doublet-dominated states have widths up to $12.5\%$. Decays into the scalar fields that set the speeds and into the radion are not included.

\section{Model B: Light Gluon, Heavy Fermions}
Reordering the hierarchy of KK mass scales, we keep every fermion and the electroweak fields at the conventional speed and slow only the gluon's fifth-direction propagation. On the same geometry, a gluon speed $u_g=0.44,0.48,0.52$ gives first KK gluons at $5.5,6,6.5\TeV$, respectively, with all other inputs fixed. The corresponding gluon coefficients $\eta_g$ in Eq.~\eqref{eq:speeds} satisfy $\eta_g\nu^2/\Lambda_0^5=0.8064,0.7696,0.7296$. Both models use the constant-mass bulk Higgs, but B retains the standard custodial representations~\cite{Agashe:2006at}. In our electroweak calculation they are needed at the model's $12.5\TeV$ electroweak KK scale. Quark partners begin at $10.74\TeV$, so partner decays of the gluon are closed. Its top couplings are $1.62\,g_s$ for both chiralities, and it is a top-philic resonance with $\Gamma/M=5.47\%$ and about $64\%$ of its quark width into $t\bar t$.

Since we have neither the freedom of three-brane models to keep wavefunctions from overlapping nor the sparse Yukawa matrices of vectorlike-chain models, flavor symmetries are less automatic, constraining the allowed Yukawa structure in general. In this model the first two left-handed doublets share a common profile and the down Yukawa is aligned, so the down-type zero-mode flavor-changing neutral currents vanish at tree level in the aligned limit, in the spirit of Ref.~\cite{Fitzpatrick:2007sa}. The meson-mixing corrections are below $3\%$ of the measured values, and the electroweak sector satisfies $\chi^2=2.88$. Here the Standard Model leptons are localized on the ultraviolet boundary, which approximates ultraviolet-localized bulk leptons. The included one-loop quark EDM estimate is below $10^{-32}\,e\,\mathrm{cm}$. Local and higher-loop CP-odd matching remains open.

The calculated $13\TeV$ resonant rate is $\sigma(pp\to G\to t\bar t)=0.110$--$0.365$~fb at $5.5\TeV$ and $0.0176$--$0.0783$~fb at $6.5\TeV$, where the range reflects the choice of parton distributions. At $5.5\TeV$ this lies a factor of two to six below the CMS limit for a $1\%$-wide resonance~\cite{CMS:2026lsc}. Because published limits depend on the signal template, a dedicated reinterpretation is needed, together with projections for the high-luminosity LHC and future colliders.

The main theoretical limitation is the color force becomes strongly coupled at a low-scale in the four-dimensional theory. This is a result of the nonrenormalizable five-dimensional theory, and a naive-dimensional-analysis estimate must be quoted at a given scale. At the first gluon resonance it is about $0.3$, but it reaches one within a few KK levels. It is about $1.55$, $1.42$ and $1.31$ at $6k_{\rm IR}$ for the $5.5$, $6$ and $6.5\TeV$ gluons, where $k_{\rm IR}=ke^{-A(L)}$, and $2.58$ at $10k_{\rm IR}$ for the $5.5\TeV$ point.

\section{Relation to Earlier Proposals} Equation~\eqref{eq:Aeff} makes the comparison with multibrane models direct. The reduced gluon five-dimensional light speed reproduces the ordering of gauge and fermion resonance scales, and with a constant speed even the first gauge mass, but not the couplings. A speed changes how fields propagate, not where they live, so the near-universal gauge couplings of the brane construction, which come from matter being absent where the light resonances live, are not reproduced. Model B instead obtains its flavor safety from imposed common profiles and alignment, and its top-philic gluon and heavy KK gravitons remain distinguishing features. Model A, light fermions with heavy gauge bosons, realizes the opposite ordering, which a matter-terminating brane cannot produce: a gauge field cannot end on a brane while charged matter continues beyond it.

We now compare to vectorlike chains. Deconstructing the interval turns each fermion tower into a chain with hopping set by $v_\Psi$ and each gauge field into a chain of gauge groups with links set by $u_a$~\cite{Arkani-Hamed:2001kyx}. Ordinary deconstructed flavor models suffer from the resulting vector resonances. The chain models have none. The ratio $v_\Psi/u_a$ interpolates between the two: equal speeds give an ordinary extra dimension, while $v_\Psi/u_a\to0$ at fixed fermion masses removes the gauge resonances. Exponential localization is the continuum of a uniform chain. The chain models' singlet-only vectorlike fermions have a counterpart in the light right-handed partners of Model A, their accidental $U(2)$ in the imposed common doublet profile of Model B, and their down-sector mixing in the protected model of the companion paper~\cite{Companion}. The differences are equally sharp. In our models the gauge resonances are suppressed rather than absent, hierarchies come from order-one geometric parameters rather than small couplings, and the quoted $W/Z/h$ partial widths imply prompt partner decays, whereas chain states can be hierarchical and long lived.

\section{Discussion}
Several directions follow naturally. More elaborate scalar configurations allow position-dependent speeds, which can more closely mimic additional branes, and a higher-dimensional origin of the gradient fields could tie the speeds to geometry. Different hierarchies are possible: other species can be slowed, several distinct speeds can coexist, and leptons, which face stronger dipole constraints, can keep the conventional speed or be treated separately. Both models call for collider studies of their distinctive signatures: light vectorlike quarks with a broad gauge structure, or a narrow top-philic gluon below heavy partners. The companion paper~\cite{Companion} also describes a protected warped model with both towers light and solvable flat realizations, and gives detailed finite-tower flavor and electroweak diagnostics.

We do not address here all issues of flavor alignment or ultraviolet sensitivity. Rather, we present an interesting possibility that could arise in a low-energy effective theory.  The ultraviolet origin and radiative stability of the unequal speeds, the large change of the scalar field across the interval, the strong coupling of the heavy KK sector, the shape of the broad resonances, and complete EDM matching remain to be understood. With the reduced Planck mass fixed and a common scale $\Lambda_0=40\TeV$ for the scalar operators, the scalar field changes across the interval by $\Delta\sigma/\Lambda_0^{3/2}=47.1$ in A and $30.4$ in B. Controlling the scalar potential over these ranges is an additional assumption~\cite{Companion}.

The central lesson is that flavor geography and KK spectroscopy are separable. The same extra dimension can generate Standard Model hierarchies while presenting different excitation scales to fermions and gauge fields. The separation works in either direction, yielding light fermion partners beneath a broad gauge structure or a narrow gauge resonance beneath heavy fermions, and in the first case it realizes an ordering that terminating matter on a brane cannot. Species-dependent lightcones thus offer a new handle on the long-standing tension between low-scale flavor and resonance physics.

\begin{acknowledgments}
We thank Rashmish Mishra for very helpful discussions and suggestions. L.R. and M.R. were supported by the Gravity, Spacetime, and Particle Physics (GRASP) Initiative at Harvard University. L.R. was also supported by the Moore Foundation Award 8342.
AI tools assisted with calculations, reference checking, and the writing of this paper.
\end{acknowledgments}

\bibliographystyle{apsrev4-2}
\bibliography{refs}
\end{document}